\documentclass[twocolumn,showpacs,preprintnumbers,amsmath,amssymb]{revtex4-1}
\usepackage{graphicx}
\usepackage{dcolumn}
\usepackage{bm}

\newcommand{\be}{\begin{eqnarray}}
\newcommand{\ee}{\end{eqnarray}}
\newcommand{\nn}{\nonumber\\}
\newcommand{\la}{\langle}
\newcommand{\ra}{\rangle}

\begin{document}


\title{Atomic Quantum Simulation of the Lattice Gauge-Higgs Model:
Higgs Couplings and Emergence of Exact Local Gauge Symmetry}

\author{Kenichi Kasamatsu$^1$, Ikuo Ichinose$^2$, and Tetsuo Matsui$^1$}
\affiliation{
$^1$Department of Physics, Kinki University, Higashi-Osaka, Osaka 577-8502, Japan\\
$^2$Department of Applied Physics, Nagoya Institute of Technology, 
Nagoya  466-8555, Japan
}

\date{\today}

\begin{abstract}
Recently, the possibility of quantum simulation of dynamical
 gauge fields was pointed out by using 
a system of cold atoms trapped on each link in an optical lattice. 
However, to implement exact local gauge invariance, fine-tuning the 
interaction parameters among atoms is necessary. 
In the present Letter, we study the effect of violation of the U(1) local gauge 
invariance by  relaxing  the fine-tuning of the parameters
and showing that
a wide variety of cold atoms 
is still to be a faithful quantum simulator for 
a U(1) gauge-Higgs model containing a Higgs field sitting on sites.
The clarification of the dynamics of this gauge-Higgs model 
sheds some light upon various unsolved problems including the 
inflation process of the early Universe. 
We study the phase structure of this model by Monte Carlo simulation, and 
also discuss the atomic characteristics of the Higgs phase in 
each simulator.
\end{abstract}

\maketitle

In the past decade, the possibility of using ultracold atoms in an optical lattice 
(OL) as a simulator for various models in quantum physics
seems to have become increasingly more realistic \cite{book,Bloch}.
In particular, one interesting possibility is to simulate lattice gauge theories
 (LGTs) by placing several kinds of cold atoms on the links of an OL  according to
certain rules \cite{Buchler,Tewari,Weimer,1,2,3,4,5,6,Wiese}.
Several proposals for pure U(1) LGTs \cite{pure}
were given in Refs.\cite{Buchler,Tewari,Weimer,1,2,3} and  
later extended to quantum 
electrodynamics with dynamical fermionic matter \cite{4,5} 
and non-Abelian gauge models \cite{6}.

Shortly after their
introduction by Wilson \cite{wilson}, LGTs have been studied
quite extensively, mainly in high-energy physics,
by using both analytical methods and  Monte Carlo (MC) simulations,
and their various properties have now been clarified.
However, the above-mentioned approach using cold atoms in an OL   
provides us with another interesting method for studying LGTs. 
As an example of expected results, the authors of 
Ref.\cite{1} refer to clarification of  dynamics of electric strings in the 
confinement phase. 
The atomic quantum simulations allow one to address problems 
which cannot be solved by conventional MC methods because 
of sign problem. 
One characteristic point of this cold-atom approach is that the
equivalence to the gauge system is established only under some specific conditions. 
For example, in Refs.\cite{1,2,3,4,5,6}, one needs to fine-tune a set of interaction
parameters; in other words, the {\em local gauge symmetry
is explicitly lost} when these parameters deviate from their optimal values.  

The above-mentioned point naturally
poses us  serious and important  questions on the stability of gauge symmetry, 
and potential subtlety of experimental results of cold atoms as 
simulators of LGTs, because the 
above conditions are generally not satisfied exactly or  easily violated 
in actual cold-atom systems. 
In this Letter, we address this problem
semi-quantitatively and exhibit the allowed range of violation
of the above conditions, such as the regime of interaction parameters,
within which the results can be regarded as having LGT properties.
In addition, we find that the cold atoms in question may be used as 
a quantum simulator
of a wide class of U(1) gauge-Higgs model, 
i.e., a Ginzburg-Landau-type model in the London limit coupled with the 
gauge field, 
the dynamics of which should offer us important insights on
several fields including inflational cosmology \cite{inflation}.

Let us start with the path-integral representation of
the partition function $Z$ of the compact U(1) pure LGT, the reference system 
of the present study: 
\be
Z&=&\int [dU] \exp(A),\ 
\int [dU]\equiv\prod_{x,\mu}\int_0^{2\pi}\frac{d\theta_{x\mu}}{2\pi},
\nn
A&=& \frac{c_2}{2}\sum_x\sum_{\mu < \nu}\bar{U}_{x\nu}\bar{U}_{x+\nu,\mu}U_{x+\mu,\nu}U_{x\mu}+{\rm c.c.}\nn
&=&c_2\sum_x\sum_{\mu < \nu}\cos\theta_{x\mu\nu},\ \theta_{x\mu\nu}\equiv\nabla_{\mu}\theta_{x\nu}-\nabla_{\nu}\theta_{x\mu},
\nn
U_{x\mu}&\equiv&\exp(i\theta_{x\mu}),\ 
\nabla_\mu f_x \equiv  f_{x+\mu}-f_x,
\label{z}
\ee
where $x =(x_1,x_2,x_3,x_4)$ 
is the site index of the 3+1=4D lattice ($x_4$ is the imaginary time in
the path-integral approach) and 
$\mu$ and $\nu$ $(=1,2,3,4)$ are the direction indices that we also use 
as the unit vectors in the $\mu$ and $\nu$-th directions. 
The angle variable $\theta_{x\mu} \in [0,2\pi)$ and its exponential $U_{x\mu}$ 
are the gauge variables defined
on the link $(x,x+\mu) $ \cite{continuum}. 
The bar in $\bar{U}_{x\mu}$ implies 
complex conjugate, and $c_2 (\equiv 1/e^2)$ 
is the inverse self-gauge-coupling constant.
The product of four $U_{x\mu}$ is 
invariant under the local ($x$-dependent) U(1) gauge transformation,
\be
U_{x\mu} \to U'_{x\mu} \equiv 
V_{x+\mu}U_{x\mu}\bar{V}_x,\ V_x\equiv \exp(i\Lambda_x),
\label{gt}
\ee
and so are the field strength $\theta_{x\mu\nu}$ and the action $A$. 
It is known \cite{4du1} that the system has a weak
first-order phase transition at $c_2 =c_{2c} \simeq 1.0$.
For $c_2 < c_{2c}$ ($> c_{2c}$) the system is in the confinement (Coulomb) phase in which
the fluctuations of $\theta_{x\mu}$ are strong (weak).
In the Coulomb phase,  $\theta_{x\mu}$ describes
almost-free massless particles, which correspond to 
photons in electromagnetism \cite{continuum}.

To obtain the quantum Hamiltonian $\hat{H}$ for $Z$, 
let us focus on the space-time plaquette term 
$\cos \theta_{xi4}$ in $Z$ with the spatial direction index
$i (=1,2,3)$ and rewrite it as 
\be
&&\exp\left(c_2\cos\theta_{xi4}\right)
\simeq\sum_{m_{xi}\in{\bf Z}}
\exp\left[-\frac{c_2}{2}(\theta_{xi4}-2\pi m_{xi})^2\right]
\nn
&\propto&\sum_{E_{xi} \in {\bf Z}}\exp\left[-iE_{xi}(\nabla_i\theta_{x4}-
\nabla_4\theta_{xi})
-\frac{1}{2c_2}E_{xi}^2\right],
\label{villain}
\ee
where we used the Villain (periodic Gaussian) approximation in the first
line and Poisson's summation formula in the second line. 
The term $iE_{xi}\nabla_4\theta_{xi}\simeq 
id\tau E_{xi}\dot{\theta}_{xi}$
($\tau$ is the imaginary time and 
$\dot{f}\equiv df/d\tau$) shows that the 
integer-valued field $E_{xi}$ on the spatial link $(x,x+i)$ is the conjugate
momentum of $\theta_{xi}$. Thus, the corresponding operators
at spatial site $r=(x_1,x_2,x_3)$ satisfy the canonical commutation relation
$[\hat{E}_{ri},\hat{\theta}_{r'i'}]=-i\delta_{rr'}\delta_{ii'}$.
The operator $\hat{E}_{ri}$ represents the electric field in electromagnetism but
has integer eigenvalues owing to the compactness (periodicity) of
$A$ under $\theta_{x\mu}\to\theta_{x\mu}+2\pi$. 
The integration over $\theta_{x4}$ can be  performed as
\be
G&\equiv&\int \prod_x d\theta_{x4}\exp(-i\sum_{x,i}E_{xi}\nabla_i
\theta_{x4})
=\prod_{x}\delta_{Q_x,0},\nn
 Q_x&\equiv&\sum_i\nabla_iE_{xi},
\label{gauss}
\ee
where we used $\sum_{x,i}E_{xi}\nabla_i\theta_{x4}=-\sum_{x,i}
\nabla_iE_{xi}\cdot\theta_{x4}$, which holds
for a lattice  with periodic boundary conditions.

One may check that the quantum Hamiltonian $\hat{H}$
which gives $Z$ 
at the inverse temperature $\beta$ 
is  just the one given by Kogut and Susskind \cite{ks},
\be 
\hspace{-0.3cm}
\hat{H}&=&\frac{1}{2c_2\Delta\tau}\sum_{r,i} \hat{E}_{ri}^2-\frac{c_2}{\Delta\tau}
\sum_{r,i<j}
\cos\hat{\theta}_{rij} 
\label{KSH}
\ee
with $\Delta\tau (\equiv \beta/N)$ being the short-time interval in 
the $\tau$ direction. 
The $\cos\hat{\theta}_{rij}$ term corresponds to the magnetic energy
$(\vec{\nabla}\times\vec{A})^2$ in the continuum \cite{continuum}.
In fact, by inserting the complete sets $\hat{1}_E=\prod_{r,i}
\sum_{E_{ri}}|\{E_{ri}\}\rangle\langle \{E_{ri}\}|$
and $\hat{1}_\theta=
 \prod_{r,i}\int d\theta_{ri}|$$\{\theta_{ri}\}\rangle\langle \{\theta_{ri}\}|$ 
in between the short-time Boltzmann factors
$\exp(-\Delta\tau \hat{H})$,
one may derive the  relations
$Z={\rm Tr}\ \hat{G}\exp(-\beta \hat{H}),\ 
\hat{G}\equiv \prod_r \delta_{\hat{Q}_r,0}$, and
$\hat{Q}_r\equiv \sum_i \nabla_i\hat{E}_{ri}$ \cite{constraint}. 
Here, $\hat{Q}_r$ is the generator of the time-independent
gauge transformation and $\hat{H}$ respects this symmetry as $[\hat{H}, 
\hat{Q}_r]=0$. The Gauss's law $\hat{Q}_r=0$ is to be imposed 
as a constraint for physical states.

Let us discuss the cold-atom studies \cite{Buchler,Tewari,Weimer,1,2,3,4,5,6,Wiese},
specifically focusing on the quantum simulator using Bose-Einstein 
condensation (BEC) in an OL \cite{1}.
We write the boson operator on the link as
$\hat{\psi}_{ri}= \sqrt{\mathstrut \hat{\rho}_{ri}}\exp[(-)^ri\hat{\theta}_{ri}],$
$(-)^r=(-)^{x_1+x_2+x_3}$, where we use the same letter $\theta_{ri}$ as $\theta_{x\mu}$ in Eq.(\ref{z})
because the former is to be identified as the latter.
We start with the following atomic Hamiltonian \cite{1}, 
\be
\hat{H}_{\rm a}\!=\!\sum_{r,a,b}\!\left[g_{ab}\hat{\rho}_{ra}\hat{\rho}_{rb}
\!+\frac{V_0}{2}\hat{\rho}_{ra}^2\!+g'_{ab}(\hat{\psi}^\dagger_{ra}\hat
{\psi}_{rb}+\mbox{H.c.})\right],
\label{atomich}
\ee 
where $a,b=1\sim 6$ counts the links emanating from each site.
The $g_{ab}$-term describes the densty-density interaction, 
the $V_0 (>0)$-term is the on-link repulsion, and the $g'_{ab}$-term
is the hopping term induced by external electromagnetic fields.
We assume that the average  $\la \hat{\rho}_{ri}\ra = \rho_0$ is 
homogeneous and large, $\rho_0 \gg 1$, and set $\hat{\rho}_{ri}=
\rho_0+(-)^r\hat{\eta}_{ri}$, where $\hat{\eta}_{ri}$
is the density fluctuation. 
Then, by choosing $g_{ab}$ and $g'_{ab}$ suitably \cite{1,2,3,4,5,6} as 
$g_{ab} = g\ (> 0)$ for any $a$ and $b$, 
$g'_{ab}\simeq 0$ for parallel link pairs, and
$g'_{ab}= g'$ for perpendicular pairs, 
$\hat{H}_{\rm a}$ is rewritten effectively as  
\be
\hspace{-0.5cm}
\hat{H}_{\rm a}&\!=&\!\frac{1}{2\gamma^2}\sum_r
\Big(\sum_i\nabla_i\hat{\eta}_{ri}\Big)^2\!+\!V_0\sum_{r,i}\hat{\eta}_{ri}^2
\!+\hat{H}_{\rm L}(\{\hat{\theta}_{ri}\}),\nn
\hat{H}_{\rm L} &=& 
2g'\rho_0\sum_{r, i <j}\left(\cos(\hat{\theta}_{ri}-\hat{\theta}_{rj})
+\cdots\right),\
\label{hatom}
\ee
The term with $\gamma^2\ (\equiv g^{-1})$ comes from the $g_{ab}$-term and 
represents the strength of the correlation 
of fluctuations $\hat{\eta}_{ra}$ around each site
(partial conservation of atomic number). 
We note that setting $g_{ab}$ independent of $a,b$
and controlling its magnitude $g=\gamma^{-2}$  may be achieved  
by designing the OL suitably or by using interspecies Feshbach resonances
\cite{1,2,3,4,5,6}. Some theoretical
ideas for the latter are also proposed \cite{zpu}. 
$\hat{H}_{\rm L}$ describes the phase correlation between  
the L-shaped nearest-neighbor (NN) links 
[the omitted terms in the sum are explicitly written in $A_{\rm L}$
of Eq.~(\ref{za}) below].
We use the coherent state $|\{\psi_{ri}\}\rangle$ 
and $\hat{1}$$ =\prod_{r,i}$$\int d\rho_{ri}$$
d\theta_{ri}$ $|\{\psi_{ri}\}\rangle$$\langle\{\psi_{ri}\}|$ to
obtain the path-integral for 
$Z_{\rm a}$$={\rm Tr}$$\ \exp(-\beta\hat{H}_{\rm a})$ as
\be
\hspace{-0.1cm}
Z_{\rm a}
&=&\!\!
\int\prod_{x,i}[d\eta_{xi}d\theta_{xi}]
\exp\Big[\sum_{x,i}\!\Big(\!-i\eta_{xi}
\nabla_4\theta_{xi}\!-\Delta\tau V_0 \eta_{xi}^2 \Big) \nn
&&-\frac{\Delta\tau}{2\gamma^2}\sum_x\Big(\sum_i\nabla_i\eta_{xi}\Big)^2\!
-\!\Delta\tau\sum_{x_4}H_{\rm L}(\{\theta_{xi}\})\Big].
\label{eq1}
\ee
The first term in the exponent in R.H.S. 
comes from
$\sum_{x_4}\bar{\psi}_{xi}\nabla_4\psi_{xi}\simeq 
i\sum_{x_4}\eta_{xi}\nabla_4{\theta}_{xi}$ 
and shows that $-\hat{\eta}_{ri}$ is the conjugate momentum
of $\hat{\theta}_{ri}$, whereby $\hat{E}_{ri}=-\hat{\eta}_{ri}$. 

The Gaussian factor 
$\tilde{G} \equiv \prod_x \exp[ (-\Delta \tau /2 \gamma^2) Q_x^2]$
in Eq.~(\ref{eq1}) with $Q_x\equiv-\sum_i\nabla_i\eta_{xi}$ 
shows that the Gauss's law $Q_x=0$ of Eq.~(\ref{gauss}) is 
achieved by $\tilde{G}\propto \prod_x\delta(Q_x)$ only at $\gamma \to 0$,
and it is now shifted  for $\gamma > 0$ to a Gaussian distribution with
$Q_x^2 \lesssim \gamma^2/\Delta\tau$. Thus, $\gamma$ 
is a parameter used to measure the violation of Gauss's law.
Note that $\tilde{G}$ may be written as 
\be
\tilde{G} \simeq \!
\int_0^{2\pi}\! \prod_{x}\! \frac{d\theta_{x4}}{2\pi} 
\exp\! \left(\! \frac{\gamma^2}{\Delta\tau}\cos\theta_{x4}
-i \theta_{x4}\sum_i\!\nabla_i\eta_{xi}\! \right).
\label{gauss3}
\ee
By integrating Eq.~(\ref{eq1}) with Eq.~(\ref{gauss3}) over 
$\eta_{xi}\in (-\infty,\infty)$, 
one  obtains a term $-(4\Delta \tau V_0)^{-1}
(\nabla_4{\theta}_{xi}-\nabla_i\theta_{x4})^2$,
which is a part of Gaussian Maxwell term.
However, 
this result should be improved to respect 
the periodicity under $\theta_{xi}\to\theta_{xi}+2\pi$,
because $\theta_{xi}$ is the phase of the condensate.
This Gaussian term is to be replaced, e.g.,  by a periodic Gaussian
form or by the corresponding cosine form $\cos\theta_{xi4}$ as in 
Eq.~(\ref{villain})
(which may be achieved by {\it summing over the integer} $\eta_{xi}$). 
After the summation over $\eta_{xi}$,
$Z_{\rm a}$ may be expressed by the following general form;
\be
Z_{\rm a} &=&\int [dU] \exp(A_{\rm a}),\ 
A_{\rm a} = A_{\rm I}+A_{\rm P}+A_{\rm L},\nn
A_{\rm I} &=& \sum_{x,\mu}c_{1\mu} \cos\theta_{x\mu},\
A_{\rm P} =\sum_{x,\mu<\nu}c_{2\mu\nu}\cos\theta_{x\mu\nu},\nn
A_{\rm L} &=& \sum_{x,\mu <\nu}c_{3\mu\nu}\Big[
\cos(\theta_{x\mu}-\theta_{x\nu})+\cos(\theta_{x\mu}+\theta_{x+\mu,\nu})
\nn
&+&
\cos(\theta_{x+\mu,\nu}-\theta_{x+\nu,\mu})+
\cos(\theta_{x\nu}+\theta_{x+\nu,\mu})
\Big].
\label{za}
\ee

The anisotropic parameters in $A_{\rm a}$ are given as follows;
$c_{14}=\gamma^2/\Delta\tau, c_{1i}=0$ 
and $c_{2i4}\simeq (2\Delta\tau V_0)^{-1}$.
$\hat{H}_{\rm L}$ with general values of $g'$ directly
gives rise to the $A_{\rm L}$ term with $c_{3i4}=0$ and $c_{3ij}=
2g'\rho_0\Delta\tau$, while $c_{2ij}=0$ \cite{parallel}.
We note that, for $g'$ much smaller than $\gamma^{-2}$ and/or $V_0$, 
one may treat $\hat{H}_{\rm L}$ as a perturbation.
In Refs.~\cite{1,2,5}, the case $\gamma \simeq 0$ is considered 
to enforce the Gauss's law, 
and the second-order perturbation theory 
is invoked to obtain an anisotropic version of the 
Kogut-Susskind Hamiltonian (\ref{KSH}) 
as an effective Hamiltonian for the gauge-invariant subspace.
This implies $c_{2ij}\simeq \gamma^2\rho_0^2 {g'}^2\Delta \tau$ and 
$c_{3\mu\nu}=0$ in Eq.~(\ref{za}). We refer to this case later as the 
$\gamma \simeq 0$ case.

Concerning to $c_{1i}$, we note that nonvanishing 
$c_{1i}$ terms may be incorporated into the cold-atom system
by an idea discussed in Ref.~\cite{c1};
one may couple to $\hat{\psi}_{ri}$ the atomic field $\hat{a}_{ri}$ 
in another hyperfine state held in a different trapping potential
via the interaction $\hat{H}_{a\psi}=\kappa \sum_{ri}\hat{a}^\dag_{ri}
\hat{\psi}_{ri}$+H.c. If $\hat{a}_{ri}$ condenses uniformly at sufficiently high
temperatures, $\hat{a}_{ri}$ works as a BEC reservoir and 
$\hat{H}_{a\psi}$  supplies the $c_{1i}$ term effectively with
$c_{1i}=2\kappa|\la a_{ri} \ra|\sqrt{\mathstrut{\rho_0}}\Delta \tau$.
A similar idea is also discussed in Ref.~\cite{Buchler} to
generate the $c_{2ij}$ (spatial plaquette) term. 

The $A_{\rm I}$ and $A_{\rm L}$ terms in Eq.~(\ref{za}) apparently break U(1) gauge 
invariance. However, the model $Z_{\rm a}$ of Eq.~(\ref{za}) 
with general set of parameters is equivalent to
{\it another LGT with exact U(1) gauge invariance},
i.e., the U(1) gauge-Higgs model containing a Higgs field
$\phi_x$.  $\phi_x$ is a complex field defined on site $x$ and 
takes the  form $\phi_x=\exp(i\varphi_x)$, that is 
its radial excitation is frozen (so-called London limit).
The partition function of the U(1) gauge-Higgs model
$Z_{\rm GH} (= Z_{\rm a})$ is defined by
\be
Z_{\rm GH}&=& \int[d\phi][dU]\exp A_{\rm GH}(\{U_{x\mu}\},\{\phi_x\}),\nn
A_{\rm GH}&=&A'_{\rm I}+A_{\rm P}+A'_{\rm L},\ 
\int [d\phi]\equiv\prod_{x}\int_0^{2\pi}
\frac{d\varphi_x}{2\pi},\nn[-2pt]
A'_{\rm I} &=& \sum_{x,\mu}c_{1\mu} \cos(\varphi_x+\theta_{x\mu}-\varphi_{x+\mu}),\nn
A'_{\rm L} &=& \sum_{x,\mu <\nu}c_{3\mu\nu}\Big[
\cos(\varphi_{x+\nu}+\theta_{x\mu}-\theta_{x\nu}-\varphi_{x+\mu})\nn
&&+\cos(\varphi_{x}+\theta_{x\mu}+\theta_{x+\mu,\nu}-\varphi_{x+\mu+\nu})
\nn
&&+
\cos(\varphi_{x+\mu}+\theta_{x+\mu,\nu}-\theta_{x+\nu,\mu}
-\varphi_{x+\nu})\nn
&&+\cos(\varphi_{x}+\theta_{x\nu}+\theta_{x+\nu,\mu}-\varphi_{x+\nu+\mu})
\Big].
\label{zgh}
\ee
$A_{\rm GH}$ in Eq.~(\ref{zgh}) is gauge invariant under a simultaneous transformation
of Eq.~(\ref{gt}) and 
\be
\hspace{-0.5cm}\phi_x\equiv e^{i\varphi_x} \to \phi'_x=
V_x\phi_x \hspace{2mm} (\varphi_{x} &\to& \varphi'_{x}=\varphi_x+\Lambda_x).
\label{gt2}
\ee

In fact, $Z_{\rm a}$ is nothing but the gauge-fixed version
of $Z_{\rm GH}$ with the so-called unitary gauge $\varphi_x=0$. 
In short, {\em the Higgs field $\phi_x$ represents a fictitious
charged matter field to describe
the violation of chargeless Gauss's law in ultra-cold atoms},
where the general Gauss's law with a charged field is intact. 
This relation between a gauge-invariant Higgs model and its  gauge-fixed
version in the unitary gauge holds for a general action 
$\tilde{A}(\{U_{x\mu}\})$ as
\be
\int[dU]e^{\tilde{A}(\{U_{x\mu}\})}=
\int[dU][d\phi]e^{\tilde{A}(\{\bar{\phi}_{x+\mu}U_{x\mu}\phi_x\})}.
\label{fixing}
\ee
Eq.~(\ref{fixing}) is already known in high-energy physics
where the standard U(1) gauge-Higgs model is the symmetric one, 
$c_{1\mu}=c_1, c_{2\mu\nu}=c_2, c_{3\mu\nu}=0,$
and used to discuss, e.g.,  the so-called complementarity relation between 
excitations in the confinement and Higgs phases \cite{complementarity}.
However, its relevance to the quantum atomic simulator
is quite important, 
because the relation $Z_{\rm a} = Z_{\rm GH}$ 
leads to a very interesting interpretation
that the cold-atom systems proposed
in Ref.\cite{1} and the other related models \cite{2,3,4,789} 
{\it with a general set of values of parameters}
can be used as a simulator
of a wider range of field theory, i.e.,  U(1) LGT
including the Higgs couplings. 
For example, atomic simulations of the standard U(1) gauge-Higgs model above 
certainly open a new way to understand various phenomena 
including the inflation process of the early universe \cite{inflation}
and vortex dynamics of bosonized $t$-$J$ model \cite{btj}.

\begin{figure}[ht]
\vspace{-0.6cm}
\vspace{-0.6cm}
\hspace{-0.3cm}
\includegraphics[width=1.18\linewidth]{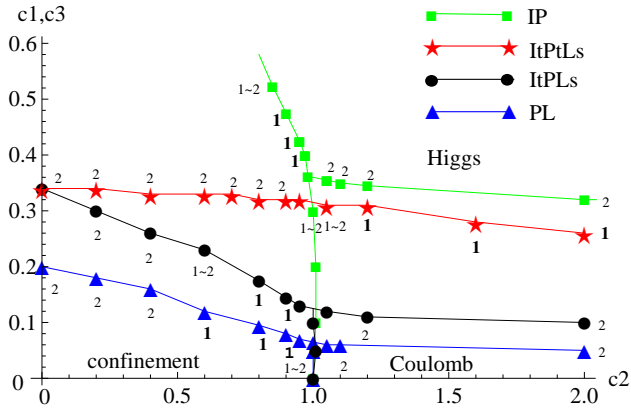}
\vspace{-1.3cm}
\caption{(Color online) Phase diagrams of the four models (\ref{higgsil}) 
in the $c_2-c_{1,3}$ plane determined by $U=\la A \ra$ and 
$C=\la A^2 \ra-\la A \ra^2$ calculated by MC simulations for a lattice 
size of $16^4$ \cite{supplement}. The vertical axis
is $c_1$ for Model IP, $c_3$ for Model PL, and
$c_1=c_3$ for Models ItPtLs and ItPLs. 
The confinement-Coulomb transition is missing in Model ItPtLs.
The number (1, 2) at each critical point 
indicates its order of transition.
The confinement-Higgs line of
Model IP terminates at $c_2 \sim 0.8$.}
\label{phasediagram}
\end{figure}

\begin{figure}[ht]
\vspace{-0.2cm}
\centering
\hspace{1cm}
\includegraphics[width=0.75\linewidth]{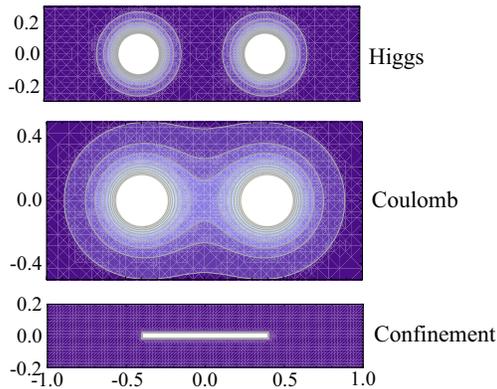}
\vspace{-0.2cm}
\caption{(Color online) Contour plot of the deviation of typical atomic density 
$\Delta \rho_{r}\equiv
(\sum_i \eta^2_{ri}/3)^{1/2}$ 
in the $x_1-x_2$ plane at  $x_3=0$ with  external sources 
of atoms $\Delta\rho_{\rm ext}=\pm \rho_1$ placed on the links emanating from 
$r=r_{\pm}=(\pm 0.4,0,0)$.
The white regions have $\Delta \rho_r$ greater than a certain value 
and the darker regions have lower $\Delta \rho_r$
The atomic density on  the link 
$(r,r+i)$  is given by $\rho_{ri}=\rho_0+\eta_{ri}$ (here we discard
the factor $(-)^r$ in front of $\eta_{ri}$ for simplicity), and 
the deviation $\eta_{xi}$ is calculated by using
the electric field $E_{ri}(=-\eta_{ri})$ 
 with a pair of external sources 
$q=\pm1$ at $r=r_{\pm}$.
In the Higgs phase, $\Delta \rho_{r}$ decreases rapidly away from the sources.
In the confinement phase, the deviation propagates from one source to the other
along a one-dimensional string (electric flux).    
}
\vspace{-0.2cm}
\label{atomicdensity}
\end{figure}

Let us study the global phase structure of the gauge-Higgs model $Z_{\rm GH}$.
We consider the following Models of $Z_{\rm GH}$ for definiteness:
\be
&&\begin{tabular}{ccccccccl}
Model&Symbol&$c_{14}$&$c_{1i}$&$c_{2i4}$&$c_{2ij}$&$c_{3i4}$&$c_{3ij}$\\
\hline
IP&{\tiny $\blacksquare$}&$c_1$&$c_1$&$c_2$&$c_2$&0&0&\\
ItPtLs&$\star$&$c_1$&$0$&$c_2$&0&0&$c_3$&\\
ItPLs&$\bullet$&$c_1$&$0$&$c_2$&$c_2$&0&$c_3$&\\
PL&{\small $\blacktriangle $}&0&0&$c_2$&$c_2$&$c_3$&$c_3$&\\
\end{tabular}
\label{higgsil}
\ee
Model ItPtLs (t denotes time and s denotes space) corresponds to 
the choice explained below Eq.~(\ref{za}).
Model PL with $c_3 =0$ corresponds to the $\gamma \simeq 0$ case, 
i.e., the pure gauge theory (\ref{KSH}) \cite{anisotropy,smallc2}.  
Figure \ref{phasediagram} shows the phase diagrams of the four Models 
in Eq.(\ref{higgsil}) in the $c_2$-$c_{1,3}$ plane obtained by standard MC 
simulations \cite{supplement}. 
There are generally three phases---Higgs, Coulomb, and  confinement---
in the order of increasing size of fluctuations of the gauge field $\theta_{x\mu}$.
These three phases can be 
characterized by the potential energy $V(r)$ stored between
two static charges with opposite signs and separated by a distance $r$, as
$V(r) \propto 1/r\ ({\rm Coulomb}),\ \exp(-mr)/r\ ({\rm Higgs}),\ 
r\ ({\rm confinement})$. 
One may distinguish each phase  in the cold atom experiments by measuring
atomic density (See Fig.\ref{atomicdensity}).      

Figure \ref{phasediagram} also 
shows that the confinement and Coulomb phases of the pure gauge theory 
along the $c_2$ axis survive only up to the phase boundary
$c_{1(3)} =c_{1(3)c}(c_2)$ (except for $c_2 \alt 0.8 $ in Model IP); 
beyond this value of $c_{1(3)}$ the system enters into a new phase, 
the Higgs phase, in which both $\theta_{x\mu}$
and $\varphi_x$ are stable. 
The expectation that the cold atoms may simulate the pure
gauge theory \cite{1,2,5} 
is assured qualitatively and globally as long
as both systems are in the same phase. This occurs for the
\\

\noindent
atomic parameters 
satisfying $c_{1(3)}< c_{1(3)c}(c_2)$.

The confinement-Coulomb transition exists only for Models 
having $c_{2i4}\neq 0$ and $c_{2ij}\neq 0$;
Model ItPtLs ($c_{2ij} =0$) has no Coulomb phase.
This is consistent with the results of pure U(1) gauge theory 
that the confinement-Coulomb transition exists for 4D system \cite{4du1} 
but not in the 3D system \cite{2d}.
For sufficiently large $c_{2i4}$ and $c_{2ij}$, $\theta_{x\mu}$ is almost frozen
$\theta_{x\mu} \simeq 0$ up to
gauge transformation and the system reduces to the XY model
with the XY spin $\phi_x=\exp(i\varphi_x)$. 
Then, the $c_{1\mu}$ term becomes the 
NN spin-interaction, $c_{1\mu}\bar{\phi}_{x+\mu}\phi_x$,
and the $c_{3\mu\nu}$ term becomes the next-NN one, 
$c_{3\mu\nu}\bar{\phi}_{x+\mu+\nu}\phi_{x}$. 
These (extended) XY models exhibit
a second-order transition both for  3D and 4D couplings,
which corresponds to the Higgs-Coulomb transition in Fig.~\ref{phasediagram}.
For small $c_{2\mu\nu}$, the confinement-Higgs
transition is missing in Model IP ($0 \leq c_2 \alt 0.8$), reflecting that
$\theta_{x\mu}$ are decoupled at $c_2=0$ 
\cite{complementarity}.
In contrast, in the other three Models, the $c_3$ term survives,
couples another set of XY spins $\exp(i\theta_{x\mu})$ on NN links, 
and gives rise to second-order transitions of the XY-model type at $c_2\simeq 0$. 

It is quite instructive to clarify the physical meaning of the Higgs phase
of the gauge system realized in atomic quantum simulators.
In the simulator using bosons \cite{1}, the Higgs phase of the effective gauge
system is nothing but the BEC state as the phase of the bosons (i.e., the gauge boson)
is stabilized coherently.
Therefore, the Higgs-confinement transition corresponds to 
the BEC transition.
On the other hand, in Refs.~\cite{4,6},
the gauge field is expressed as $\hat{U}_{ri}\simeq
(\hat{z}^{\sigma_r}_{r+i})^\dagger\hat{z}^{\sigma_r}_r$
($\sigma_r$= 1 for even $r$ and 2 for odd $r$) 
by using the Schwinger boson $\hat{z}^\sigma_{r}$, and 
the Higgs phase corresponds to the state in which the quantum state at each link 
$(r,r+i)$ is given by a coherent superposition of the particle-number states 
such as $|0\rangle_r|1\rangle_{r+i}+|1\rangle_r|0\rangle_{r+i}$.
In the double-well potential, 
this state is realized naturally, after which the Higgs phase
of the gauge system appears easily.

This way of introducing U(1) variables \cite{4,6} 
reminds us of an approach starting with an antiferromagnet
with $s=1/2$ quantum spin at each site and 
obtaining the CP$^1$+U(1) LGT \cite{sawamura}, 
which has a Schwinger-boson (CP$^1$) variable at  each site 
describing spins and an      
{\it auxiliary but dynamical U(1) gauge variables} on each link.
Although the CP$^1$+U(1) model and the present U(1) Higgs model
are different from each other, their global phase structures are
significantly similar (See Fig.~1 of Ref.~\cite{sawamura}). 

In summary, Eq.~(\ref{zgh}) is the target LGT of 
cold-atom systems that are basically those studied in Refs.~\cite{1,2} 
but with more general values of interaction parameters and a possible
atomic reservoir \cite{Buchler,c1}.
Figure \ref{phasediagram} predicts its global phase structures.
From the discussion given in Refs.~\cite{1,2,Buchler,c1} and the 
relation (\ref{fixing}), 
it may be rather universal that many cold-atom systems with multiplet
({\it ``quantum spins"}) placed on  OL links 
have their U(1) Higgs LGT counterparts. Such 
an equivalence between  cold atoms and the U(1) gauge-Higgs model
may be refereed to as {\it  ``quantum spin-gauge Higgs correspondence"}.

\newpage
\noindent
{\large{\bf Atomic Quantum Simulation of Lattice Gauge-Higgs Model:
Higgs Couplings and emergence of exact gauge symmetry --Supplemental Material--}}
\\


In this supplemental material, 
we explain some details to obtain the phase diagram Fig.~1, in particular, 
how to locate the transition points and determine the order of 
those transitions.
For this purpose, we measure the internal energy $U=\la A \ra$ and 
the specific heat $C=\la A^2\ra -\la A\ra^2$
by MC simulations \cite{mc}. 
We use the standard Metropolis algorithm \cite{metropolis} 
with the periodic boundary condition
for the lattice of size $V=L^4$ with $L$ up to 24. 
The typical number of sweeps is $30000+3000*10 \sim 100000+10000*10$,
where the first number is for thermalization
and the second number is for measurement. 
The errors of $U$ and $C$ are estimated by the standard deviation over 
10 samples. Acceptance ratios 
in updating variables are controlled to be $0.6\sim 0.8$.
We check that
the hot start ($\theta_{x\mu}, \varphi_{x\mu}$ are chosen randomly
between $[0,2\pi]$) and 
the cold start ($\theta_{x\mu}=\varphi_{x\mu}=0$) give the same results
within error margin.
The results of $U$ and $C$ are checked also by 
(i) comparison with the high-temperature expansion up to
$O(c_i^2)$, which is valid for 
small $c_i$, and (ii) comparison at large $c_2$ with 
independent simulations with setting $U_{x\mu}=1$ which should give similar  
transition point. In addition, for Model IP in Eq.~(14), we make (iii) 
comparison with the analytic 
result at $c_2=0$ (see Ref.~[23] in the text) and (iv)
comparison with the result by Jansen et al. \cite{jersak}
in which they study the phase structure of a similar model 
(Model IP with the radial component of Higgs field $\phi_{x}$ being included). 

\begin{figure}[h]
\centering
\includegraphics[width=0.8\linewidth]{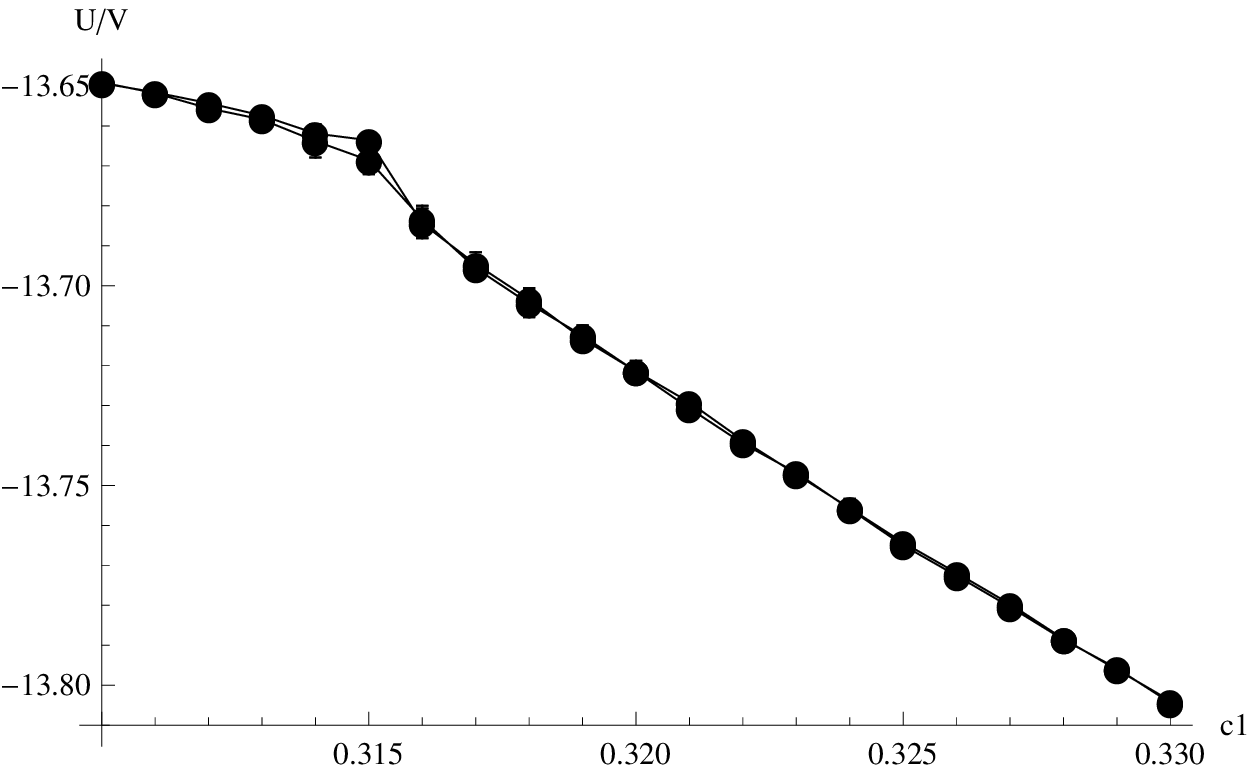}
\includegraphics[width=0.8\linewidth]{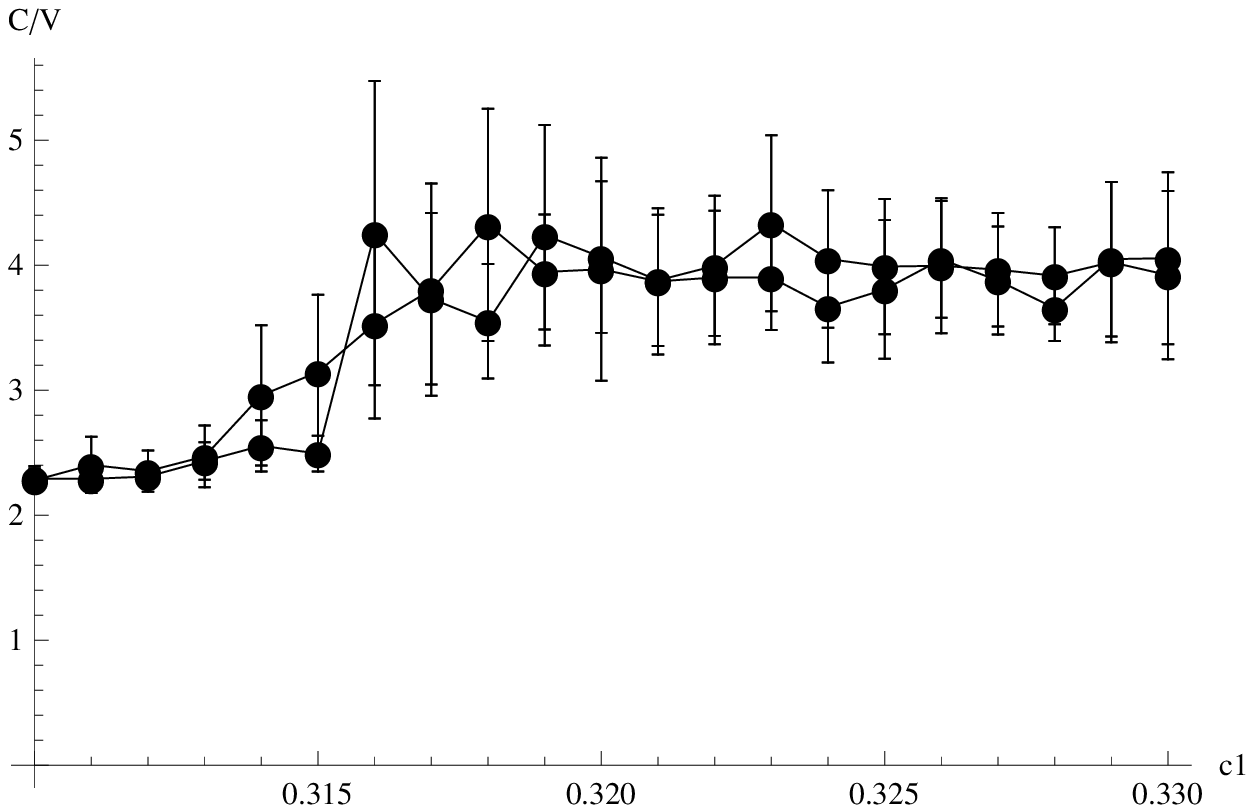}
\vspace{-0.4cm}
\caption{(Color online) 
$U/V$ and $C/V$ vs. $c_1$ for $c_2=2.5$ ($L=16$).
$c_1$. 
They indicate a second-order transition at $0.314 \alt c_1 \alt 0.322$.}
\label{c225}
\end{figure}

Let us  pick up some typical transition points for the Model IP in Eq.~(14).
Every curve of $U$ and $C$ shown below is obtained by first increasing  
the parameter $c_1$ or $c_2$ in a fixed interval with
an increment $\Delta c_{1(2)}$ and then decreasing it.
Such a go-and-back run is useful to detect a hysteresis effect.
According to their definitions in thermodynamics, 
a first-order transition has 
(i) a gap or a hysteresis loop in $U$ and (ii) a sharp peak in $C$ which usually
develops in proportional to the system size $V$, while 
a second-order transition has (i) a continuous $U$ and (ii) a gap
in $C$.   In many cases of second-order transitions,  
$C$ shows a peak which connects lower and higher-valued regions of $C$
and the peak hight develops as the system size is increased \cite{2order}.  

In Fig.~\ref{c225} we show $U$ and $C$ vs. $c_1$ for $c_2=2.5$. 
The curve $U$ itself as a function of $c_1$ is almost continuous
except for a small hysteresis loop at $c_1\sim0.315$, 
but its  derivative
with respect to $c_1$  seems to have a change (almost a gap) at 
$c_1 \sim 0.315$. 
Correspondingly, the curve $C$ globally changes
its value from the lower one around $\sim 2.2$ to the higher one around 
$\sim 4.0$  in the short interval
$0.313 \alt c_1 \alt 0.319$. 

These two behaviors accord with the 
definition of a second-order transition and therefore  we conclude
that a second-order transition takes 
place at  $0.313 \alt c_1 \alt 0.319$.
Absence of no sharp peak indicates that the associated critical exponent 
$\sigma$ is small \cite{2order}. 
We judge the hysteresis loop in $U$ is too small as an evidence
for a first-order transition.

\begin{figure}[h]
\hspace{-0.7cm}
\centering
\includegraphics[width=0.75\linewidth]{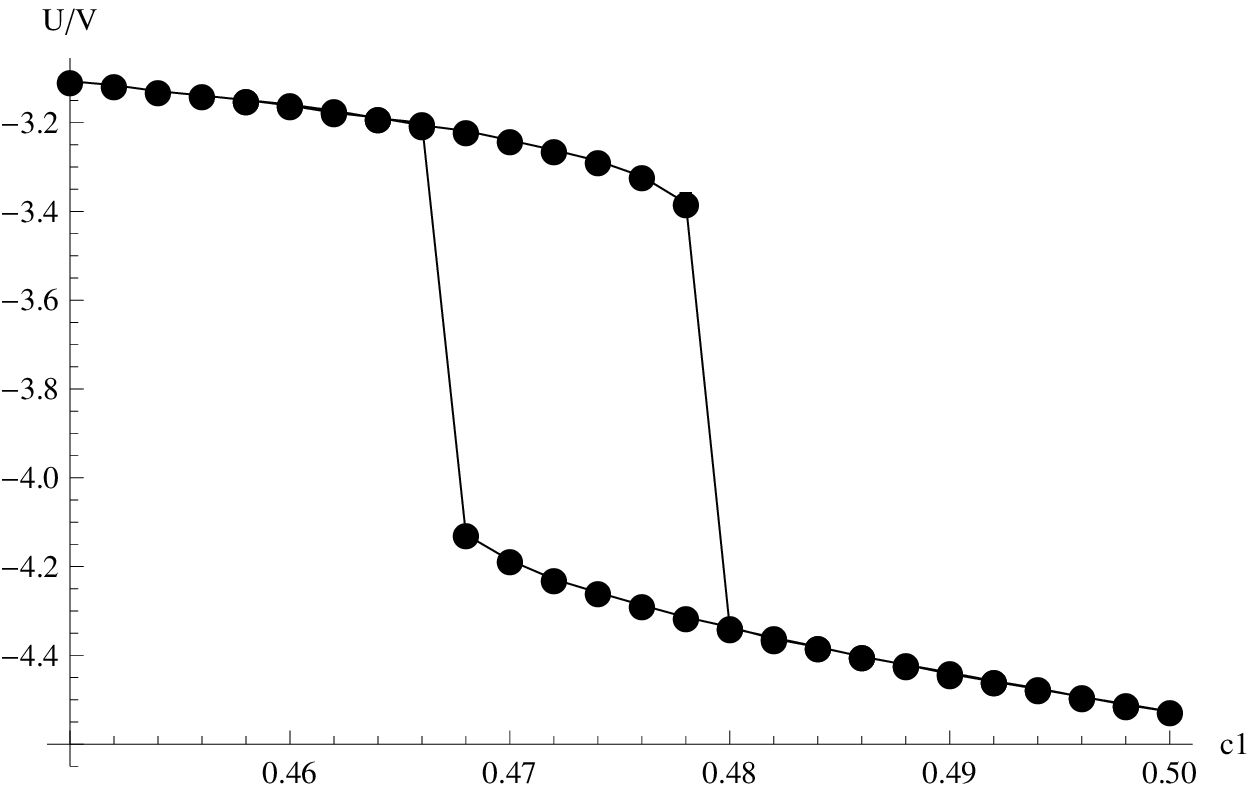}
\includegraphics[width=0.75\linewidth]{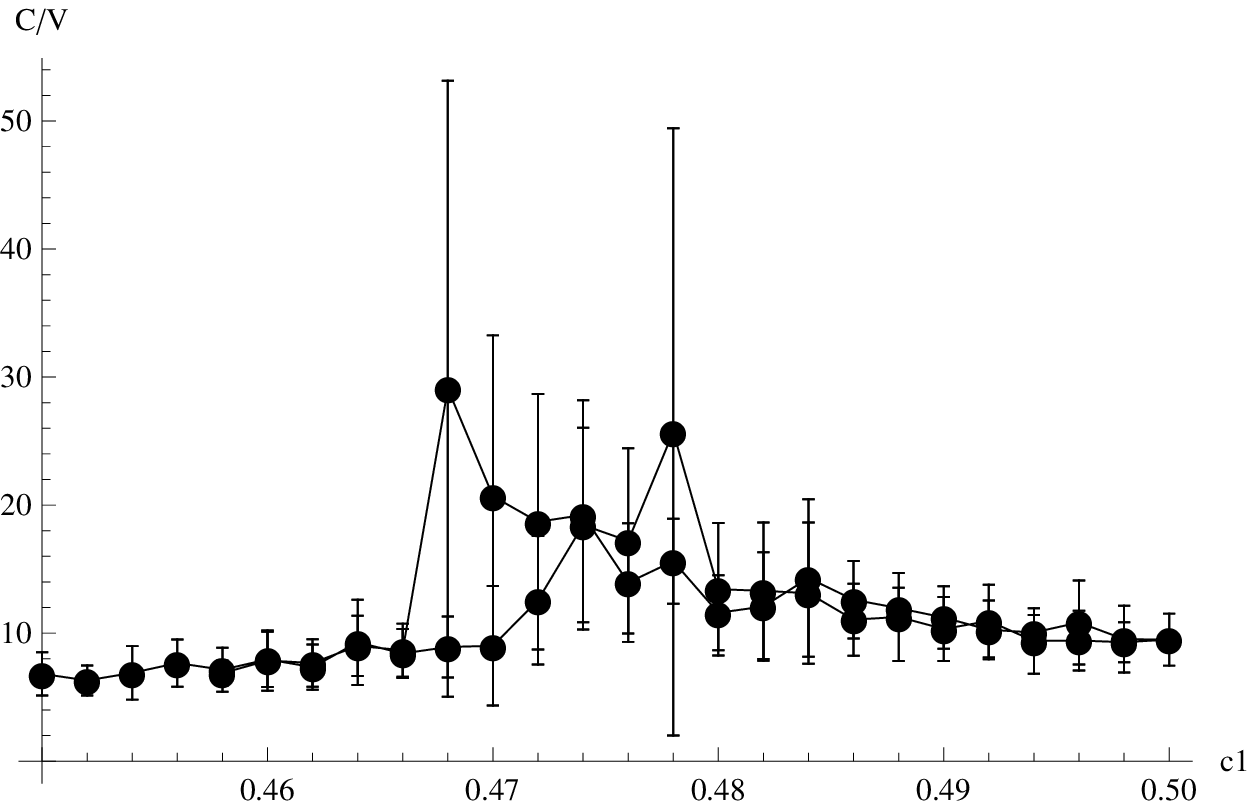}
\caption{(Color online) 
$U/V$ and $C/V$ vs. $c_1$ for $c_2=0.9$ ($L=16$).
A first-order transition takes place at $0.468 \alt c_1 \alt 0.478$.}
\label{c209}
\end{figure}
\begin{figure}[t]
\hspace{-0.7cm}
\centering
\includegraphics[width=0.75\linewidth]{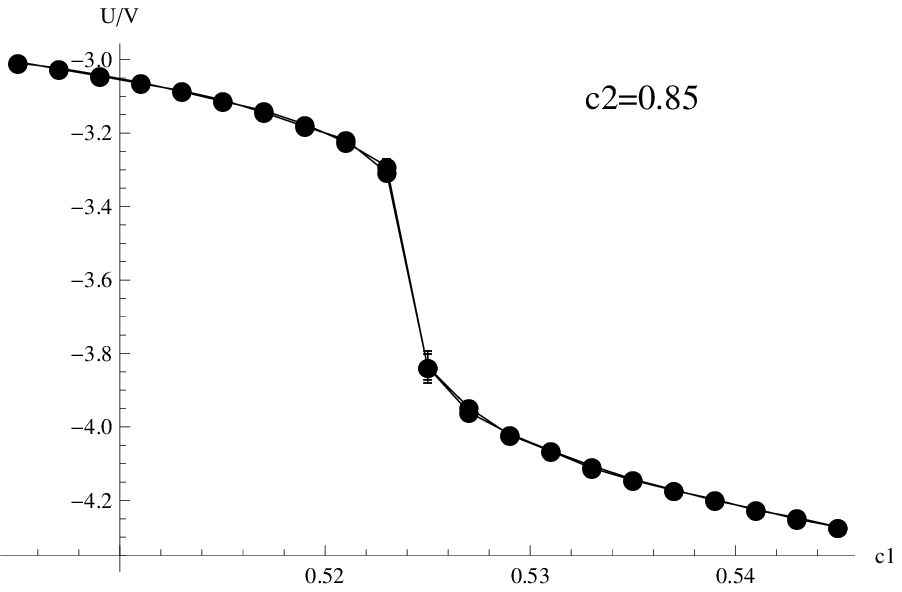}
\includegraphics[width=0.75\linewidth]{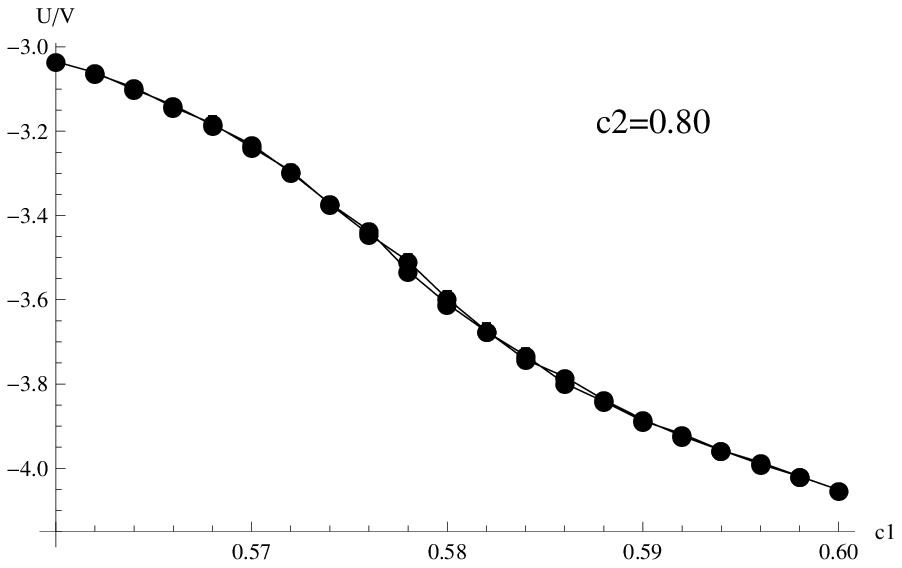}
\caption{(Color online) 
$U/V$ vs. $c_1$ for $c_2=0.85$(top) and $c_2=0.80$(bottom) ($L=16$).
For $c_2=0.85$, a weak first-order or a second-order
transition takes place at $c_1\simeq 0.524$. 
 $U/V$ for $c_2=0.80$ shows 
no jumps nor hysteresis.}
\label{c2085-80}
\end{figure}

In Fig.~\ref{c209}, we show $U$ and $C$ as a function of $c_1$ for $c_2=0.9$.
The clear hysteresis loop 
indicates the existence of a first-order transition at $0.468 \alt c_1 \alt 0.478$.
The size of 
corresponding peaks in $C$ seems not large enough
as a first-order transition, but such a phenomenon often takes place and   
is attributed to the finiteness of $\Delta c_1$. 

In Fig.~\ref{c2085-80}, we show $U$ as a function of $c_1$ for $c_2=0.85$ (top) 
and $c_2=0.80$ (bottom). For $c_2=0.85$, 
$U$ exhibits a step-function-like behavior at $c_1\simeq 0.524$, although no 
hysteresis loop appears with the present increment $\Delta c_1=0.002$.
We judge that a weak first-order or a second-order
transition takes place there. 
On the other hand, for $c_2=0.8$, $U$ looks smooth showing no
gap and hysteresis loop. Therefore we judge that no first-order transition 
takes place. Concerning to the possibility of a second-order transition,
we check whether the peak of $C$ at $c_1\simeq 0.58$
develops as the system size $L$ is increased \cite{2order}.
Our preliminary analysis using $L=20, 24$ shows that the size-dependence
is rather weak, although  the errors in $C$ are too large to draw 
a definitive conclusion. 
For a lower value $c_2=0.75$, $U$ and $C$ is smoother, and in particular,
$C$ spreads wider than 
$c_2=0.80$. From these observation,
we conclude that the line of transition should terminate at $0.75 \alt c_2 
\alt  0.85$. 
\begin{figure}[h]
\centering
\includegraphics[width=0.75\linewidth]{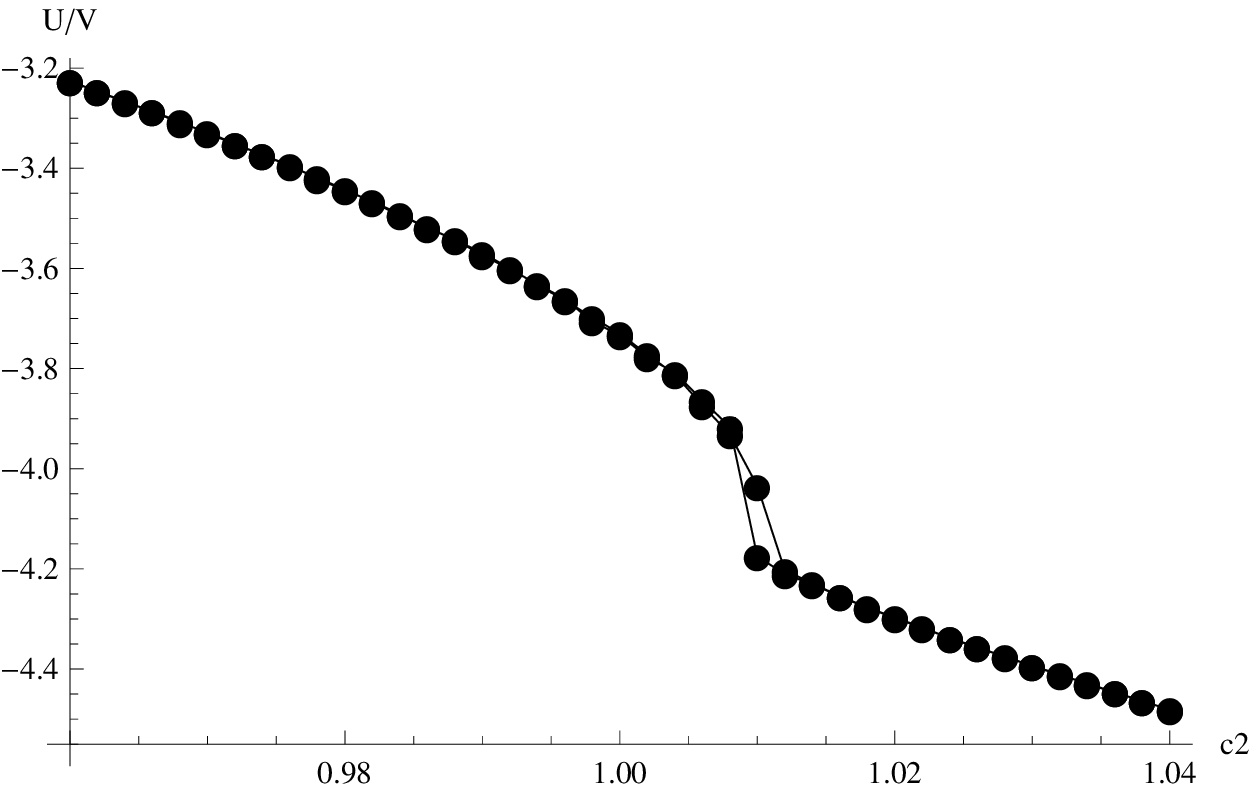}
\includegraphics[width=0.75\linewidth]{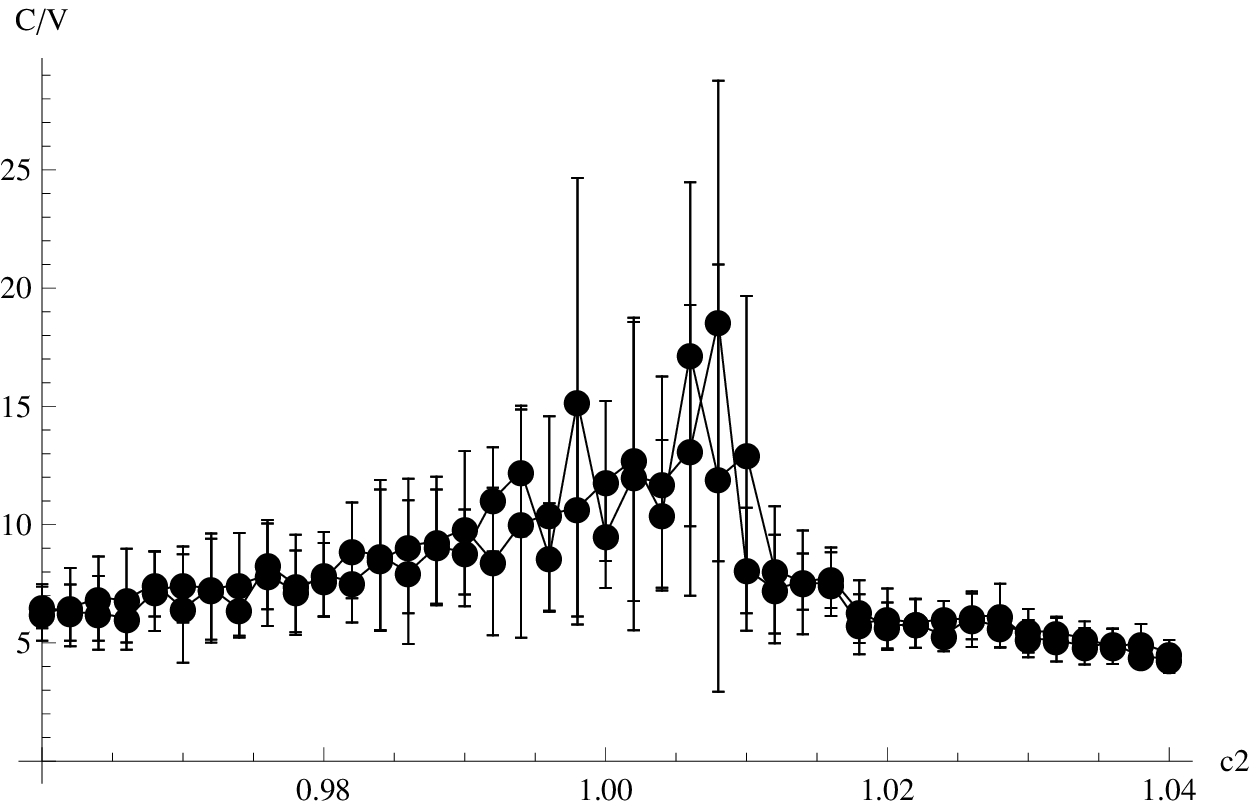}
\vspace{-0.4cm}
\caption{(Color online) 
$U/V$ and $C/V$ vs. $c_2$ for $c_1=0.3$($L=16$).
There is a weak first-order or a second-order transition
at $c_2\simeq 1.01$.}
\label{c103}
\end{figure}

In Fig.~\ref{c103}  we show $U$ and $C$ vs. $c_2$ for $c_1=0.3$.
$U$ has two branches that meet at $c_2\sim 1.01$ with different slopes
and a small hysteresis loop. We conclude that there is a weak
first-order or a second-order transition at $c_2\simeq 1.01$.

It is certainly true that more number of sweeps and 
smaller increments, $\Delta c_i$, certainly give rise to
smaller errors in $U$ and $C$ and 
more precise determination of the location and the order of the
transition points.
However, the allowed size of errors in  the location of 
the transition points drawn in Fig.~1 in the text 
is about $\Delta c_i \simeq 0.02$, i.e., almost  same as the size of 
the marks drawn there, and therefore the accuracy of the present MC study is 
almost sufficient for the purpose to draw Fig.~1 in the text.  
On the other hand, definitive determination of the order of phase transition 
for some points requires more detailed study by the MC simulations.
We hope to report on this subject in a future publication. 




\begin{thebibliography}{99}
\bibitem{pure}
The word ``pure" implies that the system
contains only gauge fields and no other
fields such as quarks, etc. 
We use it for the system with $A_{\rm a} =A_{\rm P}$ below
in Eq.~(\ref{za}).
\bibitem{continuum}
$\theta_{x\mu}$ is related to the vector potential $A_\mu(x)$ in the continuum 
space-time as $\theta_{x\mu}=aeA_{\mu}(x)$ where $a$ is the lattice 
spacing. In the formal continuum limit $a\to 0$,
the action is reduced to $A \to -(1/4)\int d^4x F_{\mu\nu}(x)F_{\mu\nu}(x)$
with $F_{\mu\nu}\equiv \partial_\mu A_\nu-\partial_\nu A_\mu$
as it should be. The partition function $Z$ can be defined without a gauge fixing
due to the compactness $\int [dU]1=1$ in contrast with $\int_{-\infty}^\infty 
dA_{\mu}(x)=\infty$.
\bibitem{parallel}
Inclusion of the parallel hopping terms to $\hat{H}_{\rm a}$ of Eq.~(\ref{atomich}) 
such as 
$\hat{\psi}_{r+j,i}\hat{\psi}_{ri}$ (both $j=i$ and $j\neq i$) is straightforward
and brings the corresponding Higgs couplings in Eqs.~(\ref{za}) and (\ref{zgh}).
\bibitem{anisotropy}
The anisotropic pure gauge theory $c_{2i4}\neq c_{2ij}$ is expected to have
the similar global phase structure as the symmetric one $c_{2\mu\nu}=c_2$
as long as all $c_{2\mu\nu}$ are nonvanishing.
See the review by J. B. Kogut in Ref.~\cite{wilson}.
\bibitem{smallc2}
We note $c_{2ij}$ in the $\gamma \simeq 0$ case of Refs.~\cite{1,2} cannot become
arbitrarily large due to its perturbative origin.
\end{thebibliography}
\begin{thebibliography}{90}
\bibitem{2order}
According to the finite-size scaling hypothesis 
[See, e.g.,  L. P. Kadanoff, Physics {\bf 2}, 263 (1966)],
$C$ near the second-order transition point behaves for large $L$ as 
$
C(c_1,L)/V=L^{\sigma/\nu}f(L^{1/\nu}\epsilon)
\label{scaling}
$ for large $L$,
where $\epsilon=(c_1-c_{1c})/c_{1c}$ and
$c_{1c}$ is the critical point for $L \to \infty$.
$f$ is the scaling function, and $\sigma, \nu$ are
the critical exponents.
\end{thebibliography}
\end{document}